\documentclass[a4paper,11pt]{article}

\usepackage{color, colortbl}
\definecolor{Gray}{gray}{0.9}
\usepackage{graphicx}
\usepackage{todonotes}
\usepackage{url}
\usepackage{authblk}

\usepackage[section]{placeins}

\AtBeginDocument{%
  \providecommand\BibTeX{{%
    \normalfont B\kern-0.5em{\scshape i\kern-0.25em b}\kern-0.8em\TeX}}}

\title{Using Online Implicit Association Tests in Opinion Polling}

\author[1,2]{Alan F. Smeaton}
\author[1,2]{Hyowon Lee}
\author[2]{Niamh Morris}
\author[2]{David Hanley}
\affil[1]{Insight Centre for Data Analytics, Dublin City University, Glasnevin, Dublin 9, Ireland}
\affil[2]{School of Computing, Dublin City University, Glasnevin, Dublin 9, Ireland}

\date{June 2020}

\begin{document}

\maketitle

\begin{abstract}
Opinion polls have now become a very important component of society because they are now a defacto component of our daily news cycle and because their results influence governments and business in ways which are not always obvious to us. However, polling is not always accurate and there have been some really inaccurate polling results which have had major influences on the world going back to the 1930s but also as recently as just the last 3 or 4 years.
In this paper we analyse the phenomenon of socially  desirable responding (shy voters) which has emerged as one of the reasons for modern day inaccurate polling. We describe how it can be exposed through implicit association tests (IATs) and we demonstrate the shy voter effect in a small survey on opinions in Ireland towards the United Kingdom. We argue for inclusion of IATs in traditional polling and point to the fact that these can be conducted accurately online, which also allows polling to reach a larger and more diverse sample of respondents in the days of Covid-19 restrictions which restricts the opportunities for poll sampling from the general public.
\end{abstract}

\section{Introduction}

Opinion polls have stealthily become an element of society, deeply embedded in our news and in decision-making.  When there is no real news to report, the results of opinion polls will drive the content of many news cycles not only for the results they produce but how these results compare with previous polls and allowing trends to be tracked.  Opinion polls happen around political elections, consumer views on products and services, constitutional decisions, referenda and plebiscites, Their influence can be enormous \ldots they can influence stock markets, election outcomes, the success or failure of new product launches and they impact public opinion.

Polling is not new and opinion polls have been around since the 19th century and in that time their {\it modus operandi} has not changed that much.  However, polling, in its current form, does have flaws and in particular the flaws are around the difficulty of recruiting subjects who represent an unbiased sample of the population, and the emergence of socially desirable responding. This latter phenomenon has been used to partially explain the disastrous performance of polling in events like the 2016 US Presidential election, the UK EU membership referendum and others.

In this paper we summarise some of the known drawbacks with current polling and we propose a greater use of the implicit association test, a simple computer-based test which can be administered online and which reveals a subject's true feelings on a topic.

The rest of this paper is organised as follows. In the next section we provide a brief history of polling followed by an overview of traditional polling, the implicit association test and how we used traditional opinion polling and IATs in a small sample of people to evaluate their views towards the United Kingdom.  That is followed by an analysis of the results and finally a concluding section completes the paper.

\section{The History of Polling}

The history of polling is one of relatively little change over many, many decades.  The principle is quite straightforward. When we want to find the likely outcome or views of a population of individuals on topics as varied as who they would vote for in a forthcoming election, what is their view on an issue like climate change or their views on same-sex marriage then we conduct an opinion poll by asking a sample of the population and extrapolating their view to the population as a whole.  The two critical things in accurate polling are firstly that the sample of people asked in the poll is large enough and reflective of the population as a whole, and secondly that they respondents give truthful answers to the poll questions. 

The earliest polls were associated with political elections, especially Presidential elections in the United States, and for about a century, from about 1830 to 1930, polling was reliable and robust. Then, in the run-up to the 1936 US Presidential election, a large poll of over 2 million individuals predicted the incumbent President, Franklin D. Roosevelt, would lose the election.  How could such a thing happen~?  Meanwhile a much smaller poll carried out by George Gallup correctly predicted Roosevelt's re-election and the world realised that sample size in polling does not overcome data bias in the sampling.  Only after the election was it realised that the larger sampled pool had selected individuals with a strong bias towards the more affluent members of society whereas Gallup's pool was more scientifically unbiased.

The 1936 election brought to the fore the issue of ignoring bias in poll sampling and since then polling has moved into mainstream to the point where it is as a source of news on a daily basis, and an input into major decisions in government and in business.
Every day our news is peppered with the results of polls and how the results of today's published poll are a change from previous polls and thus there is a shift in public opinion on whatever is the topic of the poll.
However, polling can be flawed because it is at best only a sample of the full population and recently these flaws have been revealed.

In the early 2010s Nate Silver was an American statistician who had made a reputation for predicting the outcome of major league baseball games by aggregating data from one of the most statistics-rich sports in the world \cite{baumer2014sabermetric}.  He had created an almost cult following and his methods of combining data from multiple sources had proved to be very successful. In 2012 he decided to turn his attention to politics, specifically aggregating the polling information at national and at local levels, which were proliferating the news media during the US Presidential campaign. The night before the election, he predicted the outcome almost perfectly at state level, with Barack Obama being re-elected and defeating Mitt Romney.

For the next 4 years polls, and in particular political polls, were regarded as being infallible and by the time of the 2016 US Presidential election, polling results had become a reliable source of news and a way to track trends in public opinion.  On the night before the election polls, including Nate Silver's aggregation of polls, had predicted a comfortable victory for Hilary Clinton and as we now know, all the polls got it wrong and Donald Trump was elected.  How quickly the lessons of 1936 have been forgotten.  And it wasn't just in US Presidential elections that polling was wrong.  The United Kingdom voting to leave the European Union in the Brexit referendum also in 2016, the 2017 UK general election leading to a hung parliament, the 2015 marriage equality referendum in Ireland, in all these cases and many more, pre-voting polling got it wrong.

So can anything be done about this persistent flaw in polling for major societal events~?  In \cite{BROWNBACK201838} the authors explored the effects of socially desirable responding (SDR) also known as the shy voter syndrome, where poll respondents feel their actual feelings and views on a topic may be socially unacceptable and so they give fake answers and say that they will vote for X, whereas in fact they will vote for Y.
This is especially true of sensitive issues like racism, gender, immigration or same-sex marriage.  In \cite{BROWNBACK201838} the authors posed a number of polarising questions to their participants and used their responses to tease out their true or implicit preferences. Their results did reveal that when applied to polls for the 2016 US Presidential election, voters for Donald Trump  were more likely to be ``shy'' about admitting to vote for him than voters for Hilary Clinton.

In this paper we take this a step further, albeit on a much smaller sized sample, by exploring how an implicit association test (IAT) on a topic can reveal people's true feelings and what the correlation for this is with their responses using a traditional poll.
\section{Polling and Implicit Association Tests (IATs)}

\subsection{Traditional Polling}

Worldwide, the market research industry generates about \$49 billion in annual revenue, according to ESOMAR and BDO~\cite{FirstResearch2020} and opinion polls are an essential part of that industry.
Polling  is the ``systematic gathering, recording, and analysis of qualitative and quantitative data about issues''~\cite{Somani2016}.  Pollsters rely on statistics and mathematical models to predict the behaviour and feelings of the general public towards different matters and events.  Opinion polls’ margins of error decrease as the sample population increases, meaning that to have a valid poll,  pollsters typically have to gather a large number of participants to get a 95\% or even a 99\% confidence level.

Traditionally, a poll was considered a good poll if it reached a large sample size, with broad representation. The move to online polling makes the process of gathering participants easier.  Recently, there has been a pattern to move from live phone polling to online polling, primarily due to the fact ``no one picks up the phone anymore''~\cite{Madrigal2018}. 

Now, telephone polls are becoming more expensive to conduct which is one of the main reasons for the shift to online polling. People are now harder than ever to contact by phone due to many not answering unknown numbers. This results in the interviewer searching for longer to make contact with a potential customer.  Also, it is increasingly difficult for polling companies to find people to call because a lot of consumers no longer make use of their landlines~\cite{Weckler2018}

\begin{figure}[h]
    \centering
    \includegraphics[width=0.8\textwidth]{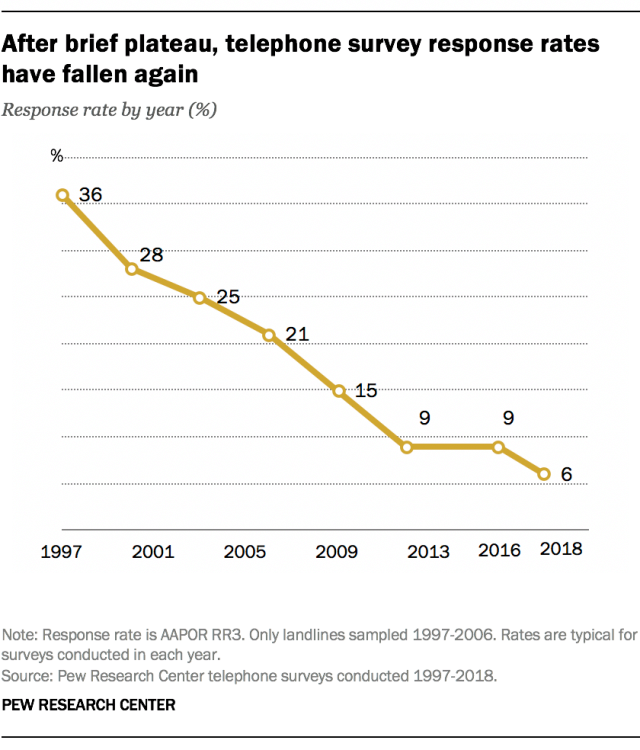}
    \caption{Decrease in Telephone Survey Responses~\cite{Kennedy2019}}
    \label{fig:Fig5}
\end{figure}

In 1997 for every 100 calls an average of 36 polls would have been conducted. If we compare that to the 6 telephone polls out of 100 calls completed in 2018 it is obvious that telephone polling is becoming untenable (see Figure~\ref{fig:Fig5}). With the decline in responses and the forever increasing costs of conducting telephone polls, this has led to the rise of online polling, and it also begs the question of who  these 6\% are !

According to Walter Shapiro of  'The New Republic’, the polling industry is in crisis, partly due to the misjudgement of the 2016 U.S. election~\cite{Shapiro2019} where Nate Silver called it ``among the greatest polling errors in primary history''.

Pollsters have been receiving criticism about the inaccuracy of their results~\cite{Dionne2003} for many years. 
These days, one criticism is that the polling industry often encounters difficulties reaching younger participants because opinion polls are seen by some of them as boring and wasteful. If young voters do agree to participate, they ''tend to hold more ideologically mixed values '' \cite{doherty2017political}. Those answers can hinder a poll's validity.

As we saw in the last section, one problem with traditional opinion poll surveys is that people can lie and give answers to polling questions that they feel will be socially accepted, e.g. ``I am not a racist''~\cite{Sagara2015}. This is known as shy voting and was observed during the Trump-Clinton presidential election in 2016~\cite{Kennedy2018} and can be found in multiple polling situations over the past few decades~\cite{Goddard2020}. Some members of the public and thus some of those included in polls were embarrassed to say they were voting for Donald Trump and so the opinion polls showed it to be a very close race but that Hillary Clinton would win the majority, whereas in fact Trump was elected~\cite{Tett2016}.

\subsection{Implicit Association Tests}

Implicit Association Tests measure the innate biases we have between concepts (e.g.Climate change) and evaluations (e.g. good or bad). The test asks a participant to categorise words that appear on the left and right sides of their screen by pressing one of two keys, typically ‘e’ for the left and ‘i’ for the right. It is important to note that the participant needs to have basic knowledge of the topic in order for the test to be valuable and also that their response is as fast as they can respond. The test is made up of five parts with each block being a series of screen presentations which elicit a response from the participant.

\begin{figure}[h]
    \centering
    \includegraphics[width=0.9\textwidth]{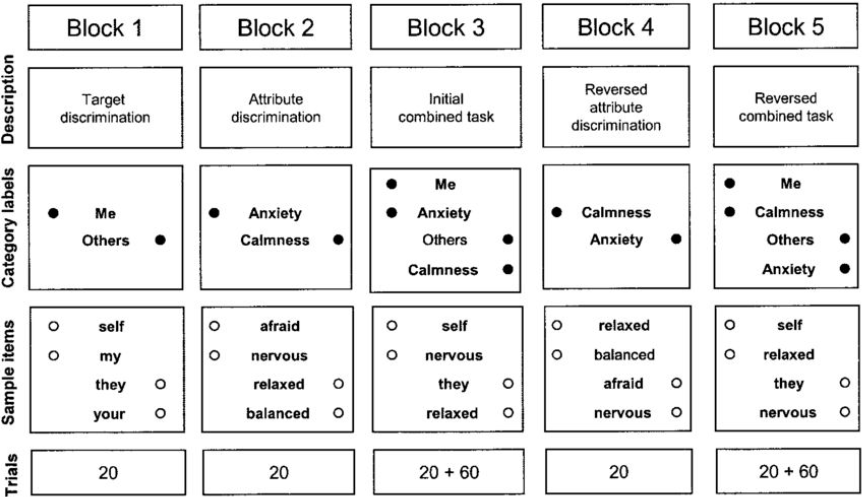}
    \caption{Example of IAT Blocks for a test on the participant's views on anxiety}
    \label{fig:AppendixA-Fig1}
\end{figure}

In the first block, the participant sorts words or images that relate to the concepts (e.g. low carbon footprint/high carbon footprint). If an energy-saving bulb was shown then the participant should select a low carbon footprint as the response. 
In the second block, the participant sorts words relating to the evaluation. If ‘pleasure’ appeared then good would be chosen. 
The third block merges the categories. e.g: low carbon or good, high carbon or bad.

\begin{figure}[h]
    \centering
    \includegraphics[width=0.7\textwidth]{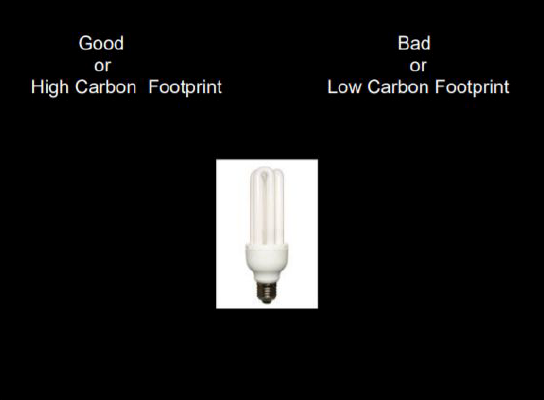}
    \caption{Participant screen from block 3, showing good or high carbon footprints versus bad or low carbon footprints (taken from Beattie and McGuire’s IAT, 2012)~\cite{Beattie2012}}
    \label{fig:AppendixA-Fig2}
\end{figure}

Then, in block 4,  the concepts switch so low carbon would be paired with bad and high carbon would be paired with good. The number of trials are increased to decrease the probability that the
participant could just have a good practice.
In the final block, the categories are placed on the opposite side to where they were previously. Low carbon would appear on the right and high carbon on the left.

The average length of time it takes the participant to categorise the words and/or images in part three versus part five, determines their test score. A person is seen to have an implicit preference for a concept if they are quicker to categorize the word when the concept is paired with good.

IATs have been considered valid for nearly twenty years. They work when conducted in person in a laboratory setting but also if administered to participants in their own homes. Carpenter {\it et al.}~\cite{Carpenter2019} verified that ``survey-software IATs appear to be reliable and valid, offer numerous advantages, and make IATs accessible for researchers who use survey software to conduct online research''.
It was also found in several  articles that ‘IAT can be just as validly administered over the Internet in participants' home environments as in a standard lab setting’ ~\cite{Houben2008} which supports the idea that IATs can be completed by participants in their own homes. This information is highly beneficial right now during this global pandemic as polls can be carried out without the need for face to face human interaction.

IATs can also measure a participant’s unconscious bias towards one of two categories e.g. gender-based roles~\cite{bias2019} such as  a nurse will be female. This is deduced by calculating the participant’s 'D-Score' (standardized difference (D) between the mean [reaction time] RT to congruent and to incongruent pairings)~\cite{Healy2015} by recording the time it takes the person to choose a category for a word and whether they categorise this correctly.

\begin{figure}[h]
    \centering
    \includegraphics[width=0.7\textwidth]{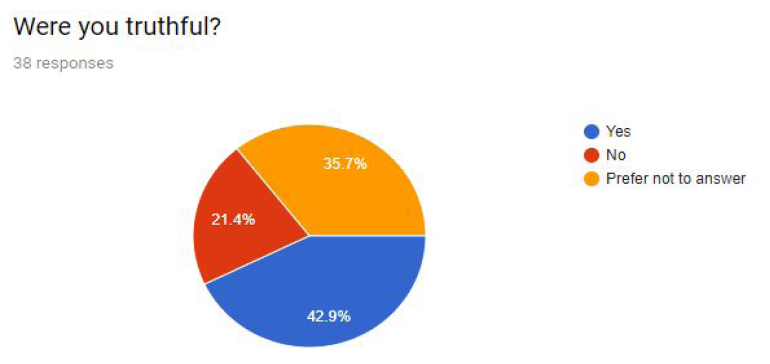}
    \caption{Responses on being asked if they were truthful when answering opinion poll}
    \label{fig:Fig8}
\end{figure}

\begin{figure}[h]
    \centering
    \includegraphics[width=0.7\textwidth]{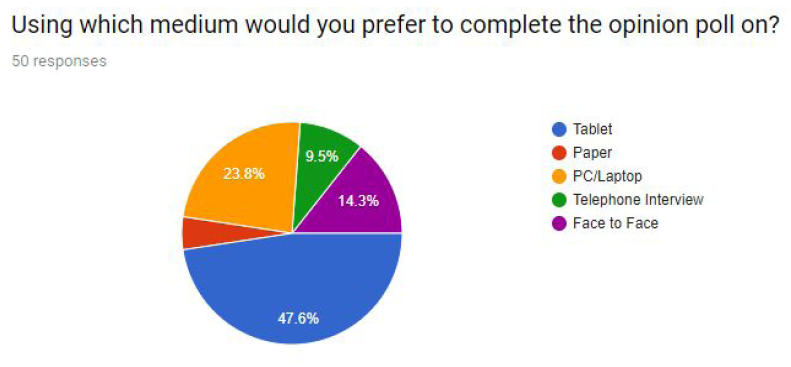}
    \caption{Participants’ opinion poll medium preferences}
    \label{fig:Fig9}
\end{figure}

To explore this ourselves, we administered a short Google form with the results shown in Figures~\ref{fig:Fig8} and \ref{fig:Fig9}).  Our sample Size was 50 and from this we found that
the percentage of participants who had previously participated in an opinion poll ws 76\%

\noindent 
The results from our questionnaire found that participants over the age of 56 gave traditional polls a rating of at least 3 out of 5 on whether they found the experience enjoyable, whereas younger participants said it was too tedious and mostly gave scores of 1 or 2.

Our main method for contacting potential participants was through the telephone and our survey  itself was, of course, flawed in that it is susceptible to exactly the kind of shy voter phenomenon we are addressing in this paper.

\subsection{IATs and Questionnaires}

To prove the value of IATs in  polling, an IAT was administered to 26 participants on Zoom. The test was created using OpenIAT and PsychoPy on the topic of `How amiable Irish people are towards the United Kingdom’. 

PsychoPy is an open-source software package written in Python, primarily used in neuroscience and experimental psychology research and we used PsychoPy to create our IAT.
Pavlovia is a website created by the PsychoPy team and it allows for the hosting and running of an implicit association test with the ability to download the results.

\begin{figure}[h]
    \centering
    \includegraphics[width=1.0\textwidth]{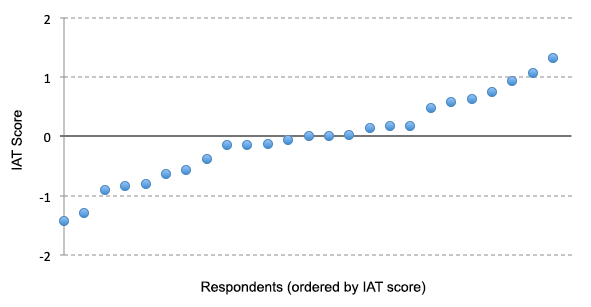}
    \caption{Distribution of 27 IAT scores}
    \label{fig:Fig11}
\end{figure}

Each participant was given a random 4 digit identification number and completed the IAT through remote access on Zoom. The gender and age bracket for each participant was recorded so that analysis could be undertaken. The results were then analysed and ranked from highest IAT score to the lowest as shown in Figure~\ref{fig:Fig11}.

What this told us was that there is almost a balanced spread of implicit biases for and against the IAT topic, with approximately the same numbers of pro, neutral and against the topic

Roughly two weeks after completing the IAT, participants were asked to complete a traditional questionnaire about their feelings towards the United Kingdom. The questionnare results for each participant were then converted to a score so that a comparison could be made between the IAT rankings and the questionnaire score rankings. A  list of the questionnaire questions and its coding scheme used for comparison is given in Table~\ref{table:Qeustionnaire}.

\begin{table}[h]
\caption{Questionnaire: questions and answer options (blue rows indicate the questions used for the correlation analysis)}
\begin{tabular}{|p{7cm}|p{5cm}|}
\hline
\rowcolor{Gray}
{\bf Question} & {\bf Answer Options (Coded)} \\ \hline
\rowcolor[HTML]{d5e1ed}
Q1. If a member of the Royal family visited your hometown, how likely would you be to go see them? & -2: Very likely, -1: Likely, 0: Neutral, 1: Less likely, 2: Not likely \\ \hline
\rowcolor[HTML]{d5e1ed}
Q2. If the UK were to go to war, would you be happy for Irish troops to join them? & -2: Yes, 2: No \\ \hline
\rowcolor[HTML]{d5e1ed}
Q3. What was your reaction to the news of Prince Charles contracting Covid-19? & -2: Sympathetic, 0:  Not care, 2: Unsympathetic  \\ \hline
\rowcolor[HTML]{d5e1ed}
Q4. Did you watch any of the Royal Weddings live on TV? &  -2: Yes,  2: No \\ \hline
\rowcolor[HTML]{d5e1ed}
Q5. Are you able to name all of Will and Kate's children? & -2: Yes, 0: Not all, 2: No \\ \hline
Q6. Do you know if Prince Philip is alive or dead? &  \\ \hline
Q7. Who is your favourite member of the Royal family? &  \\ \hline
Q8. Where is your favourite football team located? &  \\ \hline
\rowcolor[HTML]{d5e1ed}
Q9. Would you be happy for some of Ireland's emergency Personal Protective Equipment(PPE) to be shared with the UK during the Covid-19 outbreak & -2: Yes, 2: No \\ \hline
\rowcolor[HTML]{d5e1ed}
Q10. How did you feel when the UK left the EU in January 2020? & -2: Very happy, -1: Happy, 0: Didn't care, 1: Unhappy, 2: Very unhappy \\ \hline
Q11. During the financial crisis in Ireland in 2008, the UK provided substantial financial assistance to Ireland. If the UK experienced similar financial difficulty should Ireland do likewise? &  \\ \hline

\end{tabular}
\label{table:Qeustionnaire}
\end{table}

\begin{figure}[h]
    \centering
    \includegraphics[width=0.95\textwidth]{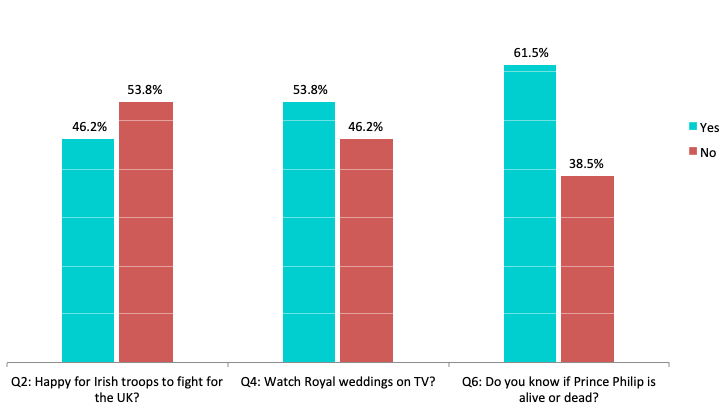}
    \caption{Results from some of the polarising questions}
    \label{fig:Fig12}
\end{figure}

In computing an overall score for each participant based on their answers in the questionnaire, the individual results  need to be weighted for different questions based on the variance of answers to each question where the higher the variance, the larger the range that exists in the answers and therefore the higher the value of a participant giving an uncommon answer. For example,  question three - \textit{What was your reaction to the news of Prince Charles contracting Covid-19?} - had the highest variance and so ansers to this were given the highest weight. As a counter-example, in question eleven - \textit{During the financial crisis in Ireland, the UK provided substantial financial assistance to Ireland. If the UK experienced similar financial difficulty should Ireland do likewise?} - the results were unanimous. This meant that there was no value in the results for the question's score because everyone felt the same. 

Each answer for each question was deemed either pro or anti the United Kingdom. For example, for \textit{If the UK were to go to war, would you be happy for Irish troops to join them?}, a participant replying 'Yes' would be more towards the UK and therefore a lower score would be given. The lower a participant’s questionnaire score, the warmer they were to the UK. A degree of polarising answers in some of the questions are shown in Figure~\ref{fig:Fig12}. Both a participant's IAT score and their questionnaire score was ranked based on other participants and a Spearman’s correlation between the two rankings was calculated. 



\section{Analysis of Results}

A total of 25 respondents completed both the IAT and  the online questionnaire, resulting in 2 sets of responses, separated in their time of gathering  by 2 weeks. We are interested to see the degree of correlation between IAT results and  questionnaire results for individual respondents. We report the process of the analysis we have performed and how we codify the questionnaire answers to represent respondents'  overall opinions towards/against UK and Ireland.  We also  describe how we remove outlier data, then explore a simple weighting scheme to differentiate amongst the questions, leading to the ranking of the respondents based on their questionnaires and their IATs. 

\textbf{Coding the answer options:} In order to calculate each respondent's level of pro-UK or pro-Ireland opinions, the answer options for each question in the questionnaire were coded in the following way: the answer options that imply an opinion supporting the UK (i.e. pro-UK) were assigned negative numbers (e.g. -2, -1), whereas the options that imply a pro-Ireland stance (i.e. pro-Ireland) were assigned positive values (e.g. 2, 1) and neutral sentiment as 0. For example, in Question 1: \textit{If a member of the royal family visited your hometown, how likely would you be to go see them~?} the answer options were \textit{-2: Very likely, -1: Likely, 0: Neutral, 1: Less likely, and 2: Not likely}, from pro-UK to pro-Ireland. 

Some questions were discarded from the analysis due to uncertainty in determining the valence, resulting in 7 questions (Q1-5, Q9 and Q10) to be used for the correlation analysis (see  Table~\ref{table:Qeustionnaire} for the questions and coded answer options). Thus a total of 7 questions and their corresponding answer options from  25 respondents were used in the analysis.

\textbf{Removing the outliers:} Once the sum of the answers from each of the questions for each respondent was calculated, it was placed against the ranking from the IAT result in order to allow direct comparison. An obvious large mismatch between the two rankings was identified. For example, respondent ID 1980 has total questionnaire score of 26, very pro-Ireland according to the coding scheme used in the study. This respondent is the 4th highest (in terms of pro-Ireland stance) amongst 25 respondents in the questionnaire ranking, but is 2nd lowest in the IAT ranking, exhibiting a very opposite result between the two methods administered, which we consider as an obvious outlier in the data. The top 4 such respondents who show extreme differences in the two methods, were identified and removed from the correlation analysis.

\textbf{Weighting for the questions:} With the codification scheme above, all 7 questions are treated as having the same importance in determining the overall opinion towards UK/Ireland of each respondent. We examined the patterns of the answers by all 25 respondents on the same questions, i.e., the way the answers were spread, distributed or polarised in some ways, as a way to determine the discriminative importance of each question.  Overall, the questions that show lower ``variance'' or higher ``polarity'' in the answers could be given higher weight so that these questions should be more prominently reflected in the questionnaire-based ranking of the respondents. In determining the weighting scheme, simple variance and reverse-deviation were tested against the overall correlation between the two rankings.

\newcolumntype{P}[1]{>{\centering\arraybackslash}p{#1}}

\begin{table}[]
\caption{Extracting the spread patterns in questionnaire answers (numbers in white cells are the number of respondents who selected the option)}
\begin{tabular}{|>{\columncolor[gray]{.9}}P{3cm}|P{0.5cm}|P{0.5cm}|P{0.5cm}|P{0.5cm}|P{0.5cm}||>{\columncolor[gray]{.9}}P{1.5cm}|>{\columncolor[gray]{.9}}P{1.5cm}|}
\hline
\rowcolor{Gray}
Question No. $\backslash$ Option & \centering -2 & -1 & 0 & 1 & 2 & Variance rank &  Reverse-Deviation rank \\ \hline
Q1 & 12 & 2 & 1 & 2 & 8 & 4 & 5 \\ \hline
Q2 & 11 & & & & 14 & 1.5 & 1.5 \\ \hline
Q3 & 20 & & 4 & & 1 & 7 & 6 \\ \hline
Q4 & 14 & & & & 11 & 1.5 & 1.5 \\ \hline
Q5 & 8 & & 4 & & 13 & 3 & 4 \\ \hline
Q9 & 16 & & & & 9 & 5 & 3 \\ \hline
Q10 & 0 & 0 & 6 & 1 & 18 & 6 & 7 \\ \hline
\end{tabular}
\label{table:VarienceTable}
\end{table}

To illustrate, Table~\ref{table:VarienceTable} shows  the number of respondents who selected each  option in the question answers (white cells), and 2 simple ranking strategies (variance and reverse-deviation). In the table, the ranking of variance in each of the questions (second last column) means that higher in the ranking (i.e. smaller the rank number) more equally distributed are the opinions of respondents amongst the available options in a question, probably becoming a less useful factor in establishing overall opinion ranking whereas lower in the ranking (i.e. larger the rank number) means that respondents' opinions were more irregular across the available options, likely suggesting a higher importance for the question. Reverse of deviation (last column) also shows a  similar ranking to variance. We experimented with these ranking metrics as  weighting schemes to differentiate the importance amongst the questions.

\begin{table}[]
\caption{Measuring the correlations between IAT and questionnaire rankings of respondents}
\begin{tabular}{|>{\columncolor[gray]{.9}}p{7cm}|P{2.3cm}|P{2.3cm}|} \hline
Weighting scheme and outliers  & Spearman Correlation & Pearson Correlation \\ \hline
No weighting - all respondents & 0.195 & 0.254 \\ \hline
Variance ranking as weighting (all respondents) & 0.197 & 0.264 \\ \hline
Variance ranking as weighting (excluding 4 outliers) & 0.429 & 0.565 \\ \hline
Reverse-deviation ranking as weighting (all respondents) & 0.307 & 0.312 \\ \hline
\rowcolor[HTML]{d5e1ed}
Reverse-deviation ranking as weighting (excluding 4 ourliers) & \textbf{0.590} & \textbf{0.624} \\ \hline
Manually-tuned weighting (Q1: 1, Q2: 0.1, Q3: 0.1, Q4: 2, Q5: 11, Q6: 1.5, Q7: 0.2) (excluding 4 ourliers) & 0.714 & 0.699 \\ \hline
\end{tabular}
\label{table:correlationtable}
\end{table}

\begin{figure}[]
    \centering
    \includegraphics[width=1.0\textwidth]{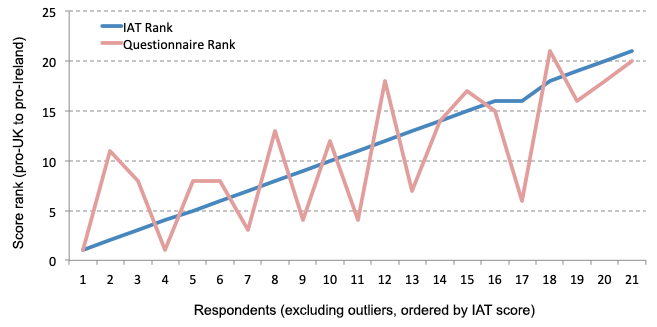}
    \caption{Comparison of participants’ IAT and questionnaire score rankings}
    \label{fig:Fig13}
\end{figure}

Table~\ref{table:correlationtable} shows the impact of weighting schemes and the removal of top 4 outlier respondents on the correlations in the results between IAT and  questionnaire rankings. As can be seen, using reverse-deviation ranking as the weighting scheme to differentiate amongst the questions and excluding the most obvious 4 outliers (second last row in the table) resulted in a Spearman correlation of 0.590 and Pearson correlation of 0.624. Using this scheme, Figure~\ref{fig:Fig13} shows the comparison of the two measures plotted by the rankings of the answers.

Manually tweaking the individual weighting factors for each question, the correlation improves as high as 0.714 (Spearman) and 0.699 (Pearson) (last row in Table~\ref{table:correlationtable}).

The weighting scheme for the individual questions could be explored  further, but here, using simple methods to take into account the answer patterns within each question by respondents, we demonstrated that a reasonably high correlation can be witnessed between IAT result and  questionnaire result.

\section{Conclusion}

In this paper we explored the relationship between traditional polling and implicit association tests as a way to expose the impact of  socially  desirable responding also known as the shy voter syndrome, in polling.

This is important because of the difficulty traditional pollsters have in reaching participants as their usual approaches to recruiting, by telephone calls, dries up and they are faced with real difficulty in ensuring the participants they use reflect a balanced spread of views. The topic is also important because the traditional approaches to polling have had some spectacular failures in the last 5 years or so, yet still we see the results of polls as major items in our news media.

We show how an implicit association test (IAT) on a topic can reveal people’s true feelings on that topic and using a small worked example we tease out Irish people's views towards the United Kingdom using both a traditional poll, and an IAT.  Even with only 25 respondents and calculating a ranking of those respondents based on their IATs and their questionnaire results, we achieve a best performance of only 0.714 for a Spearman correlation between the two.  That is achieved by  adjusting relative question weights and answer importances, in order to to maximise the correlation. The fact that we do not get a better alignment between the two rankings, by IAT and by questionnaire, tells us that even in this small sample, socially desirable responding does exist.

The paper also points to work reported from elsewhere, in addition to the work we report here, showing that IAT tests can be run successfully over the internet and that the response times for implicit association testing, measured in milliseconds, can be achieved.  The takeaway message is thus tat traditional polling has problems with shy voters and these can be addressed by using IAT tests, and these can be administered remotely.

\vspace{1cm}
\subsection*{Acknowledgements:} 

\noindent 
AS and HL would like to acknowledge support from the Insight Centre for Data Analytics which is supported by Science Foundation Ireland under Grant Number SFI/12/RC/2289\_P2
\bibliographystyle{plainurl}
\bibliography{references}

\end{document}